\documentclass[12pt,english]{article}
\usepackage{mathpazo}

\usepackage[T1]{fontenc}
\usepackage[latin9]{inputenc}
\usepackage{geometry}
\usepackage{fancyhdr}
\usepackage{wrapfig}
\usepackage{amsmath}
\usepackage{graphicx}

\makeatletter
\numberwithin{equation}{section}
\numberwithin{figure}{section}

\newcommand{\ve}[1]{\ensuremath{\mbox{\boldmath$#1$}}}

\newcommand{\vecx}{\ve {x}}
\newcommand{\Div}{\ve \nabla \cdot}
\newcommand{\Grad}{\ve  \nabla}

\usepackage{babel}

\makeatother

\usepackage{babel}
\begin{document}
\noindent \begin{flushright}
\thispagestyle{empty}
\par\end{flushright}

\begin{center}
{\LARGE \vspace{6mm}
Evaluation of the rate constant and deposition velocity for the escape
of Brownian particles over potential barriers\vspace{3mm}
}
\par\end{center}{\LARGE \par}

\begin{center}
{\large Michael W Reeks }\\
{\large{} }School of Mechanical \& Systems Engineering, \\
 University of Newcastle, UK\vspace{5mm}

\par\end{center}
\begin{abstract}
We analyze the escape of Brownian particles over potential barriers
using the Fokker-Planck equation in a similar way to that of Chandrasekhar
(Rev. Modern Phys., 1943), deriving a formula for the particle deposition
velocity to a surface as a function of the particle response time.
For very small particle response times, the particle deposition velocity
corresponds to that obtained using a quasi-steady solution of Smoluchowski's
equation and for very large particle response times, the deposition
velocity corresponds to that based on the transition state method
(E. Wigner, Trans. Faraday Soc., 1938). 
\end{abstract}

\section{Introduction}

This short paper examines the work of Chandrasekhar \cite{key-2}
on the escape of Brownian particles over potential barriers. It is
part of a classic paper that examines a number of stochastic problems
using the Fokker-Planck equation for Brownian particles. One of the
most important results of his study has been to show under what conditions
the Fokker-Planck equation transforms into an advection-gradient diffusion
equation commonly referred to as Smoluchowski's equation. This has
a significant bearing on the formula for the release rate of particles
over a potential barrier for it is generally assumed that the particle
flux out of the well can be obtained from a steady state solution
of Smoluchowski's equation. The general formula for the rate constant
for the release of particles from the potential well that Chandrasekhar
obtained as a solution of the Fokker-Planck equation, shows that the
formula based on Smoluchowski's equation is one extreme of the solution
when the spatial gradient of the forces (per unit mass of the particle)
is small compared with the particle response time which in many cases
of interest does not apply (see Reeks et al. \cite{key-4}). The fact
that Chandrasekhar obtained a relatively simple formula for the most
general result (not restricted by the value of the particle response
time) is very important to the analysis we present here.

The situation analyzed by Chandrasekhar is slightly different from
the case we wish to consider here where particles from the bulk of
the suspending fluid are subject to a potential barrier as they approach
a wall where they are finally deposited. It turns out that the analysis
is exactly the same as that used by Chandrasekhar, as indeed is the
form for the deposition flux.What is different is the way we need
to normalize the flux. In the one case when particles are trapped
in a well it is the rate constant for escape that we require by dividing
the flux by the number of particles in the well whereas in the case
of interest, it is the deposition velocity which means we divide the
current by the concentration in the bulk or more precisely when referring
to the situation in Figure \ref{fig:Potential Barrier for depositing particles}
to the concentration at the bottom of the potential barrier at $A.$

\begin{wrapfigure}{O}{0.5\columnwidth}%
\includegraphics[scale=0.37]{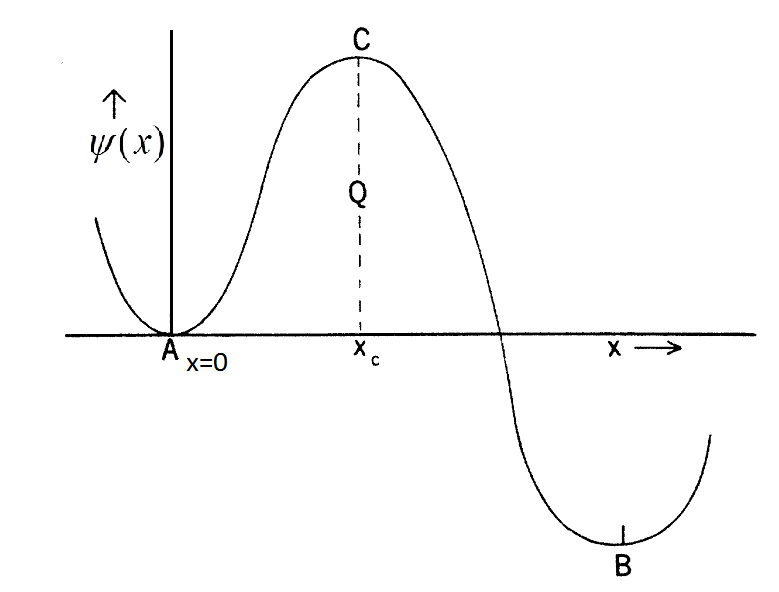}\caption{\label{fig:Potential-wells}Potential wells }
\end{wrapfigure}%

With reference to Fig (\ref{fig:Potential-wells}) we begin with the
case analyzed by Chandrasekhar \cite{key-2} where by Brownian diffusion,
particles in a 1-D potential well with minimum potential at A escape
over a potential barrier of height $Q$ at $C$ into the potential
well at B. A complete solution can be found by solving the Fokker-Planck
equation for the phase space distribution (combined spatial and velocity
distribution) for Brownian motion in a conservative field of force.
However whilst this can be achieved numerically at least for a 1 D
potential, for the case when $Q\gg$ the kinetic energy of a particle
per unit mass, analytical solutions can be found under certain realizable
conditions.

\section{Rate constant based on Smoluchowski's equation for advection diffusion. }

We refer to Chandrasekhar (1943) \cite{key-2} and the conditions
for which the Fokker-Planck equation reduces to an advection gradient
diffusion Eq.. (Smoluchowski's Eq.)

\begin{equation}
\frac{\partial w}{\partial t}=\Div\left(\frac{q}{\beta^{2}}\Grad w-\frac{\ve K}{\beta}w\right)\label{eq:}
\end{equation}
 where: 
\begin{itemize}
\item $\beta$ is the Stokes inverse particle relaxation time 
\item $w(\vecx,t)$ is the particle concentration and $\ve K(\vecx,t)$
the force acting on an individual particle both at position $\vecx,t$. 
\item $q=\beta k_{B}T/m$ is the diffusion coefficient for dispersion in
velocity space, $k_{B}$ is Boltzmann's constant and $m$ is the particle
mass and $T$ the absolute temperature. 
\end{itemize}
It is important to appreciate that Smoluchowski's equation is only
valid if we ignore any changes in time intervals $\sim\beta^{-1}$
or space intervals $\beta^{-1}\left(k_{B}T/m\right)^{1/2}$

So we may consider the escape of particles from A over the potential
barrier at C in terms of of quasi-steady concentration and a constant
(spatially and time independent) advection diffusion current $\ve j$,
implying from Eq.\ref{eq:-2} that 
\[
\ve j=q\beta^{-2}\Grad w-\beta^{-1}\ve Kw
\]
 If $\ve K(\vecx)$ is independent of $t$ and derived from a potential
$\psi(\vecx)$ such that 
\[
\ve K=-\Grad\psi
\]
 
\[
\ve j=q\beta^{-2}\Grad w-w\beta^{-1}Grad\psi
\]
 \\
 So integrating between two points $A$ and $B$ we obtain 
\begin{equation}
\ve j\cdot\int_{A}^{B}\beta exp(m\psi/k_{B}T)d\vecx=\frac{k_{B}T}{m}wexp-m\psi/k_{B}T\mid_{B}^{A}\label{eq:-1}
\end{equation}
 where we have replaced $q$ by $\beta k_{B}T/m$. Now lets determine
the rate constant or the probability per unit time that a particle
will be escape over the potential barrier at C into potential well
around B. We assume that most of the particles are in quasi equilibrium
around A and the height of the potential barrier $Q\gg k_{B}T/m$.
This means that to a high degree of approximation, a distribution
of velocities very close to Gaussian will exist in the neighborhood
of A and that the concentration $w(x)$in the neighborhood of A will
be given by 
\[
w(x)\approx w_{A\;}exp-m\psi/k_{B}T
\]
 where $w_{A}$ is the concentration of particles at A. We may assume
also that $w_{A}\gg w_{B}$ and that referring to Eq.(\ref{eq:-1})
$w\; exp-\beta q^{-1}\psi\mid_{B}^{A}\approx w_{A}$ and so from Eq.(\ref{eq:-1})
\[
j=\frac{k_{B}T}{m}w_{A}\left(\int_{A}^{B}\beta\:\: exp(m\psi/k_{B}T)dx\right)^{-1}
\]
 The value of the integral will be largely based on the value of the
integrand around $C$ where we can approximate $\psi(x)$ to 
\begin{equation}
\psi(x)\cong Q-\frac{1}{2}\omega_{C}^{2}(x-x_{c})^{2}\label{eq:potenial around C}
\end{equation}
 so 
\begin{eqnarray*}
\int_{A}^{B}exp(q\beta^{-1}\psi)dx & \cong & exp(Q/k_{B}T)\int_{-\infty}^{\infty}dx\: exp-\omega_{C}^{2}(x-x_{c})^{2}/2k_{B}T\\
 & = & exp(Q/k_{B}T)\left(\frac{2\pi k_{B}T}{m\omega_{C}^{2}}\right)^{1/2}
\end{eqnarray*}
 So we have for $j$ 
\begin{eqnarray*}
j & = & \beta\frac{k_{B}T}{m}w_{A}exp(-Q/k_{B}T)\left(\frac{2\pi k_{B}T}{m\omega_{C}^{2}}\right)^{-1/2}\\
 & = & \beta^{-1}\omega_{C}\left(\frac{k_{B}T}{2\pi m}\right)^{1/2}exp(-Q/k_{B}T)w_{A}
\end{eqnarray*}

To obtain a rate constant (probability of a particle of escaping from
the potential well in neighborhood of A into potential well at B we
need to divide $j$ by the number of particles in the well in the
neighborhood of A which we can approximate to 
\begin{eqnarray}
N_{A} & = & w_{A}\int_{-\infty}^{\infty}exp(-\frac{1}{2}m\omega_{A}^{2}x^{2}/k_{B}T)dx\nonumber \\
 & = & \left(\frac{w_{A}}{\omega_{A}}\right)\left(\frac{2\pi k_{B}T}{m}\right)^{1/2}\label{eq:-2}
\end{eqnarray}
 So finally the rate constant for the release of particles from the
potential well at A into the potential well at B is given by 
\begin{equation}
P=j/N_{A}=\left(\frac{\omega_{A}\omega_{C}}{2\pi\beta}\right)exp(-mQ/k_{B}T)\label{eq:Smoluchowski rate constant}
\end{equation}

\section{Rate constant based on the transition state method}

We refer to the case where Smoluchowski's equation does not apply
and in particular the case referring to the potential in Fig (\ref{fig:Potential-wells})
when $\omega_{C}/\beta\gg1$. In other words the particle does not
respond at all to any changes that might occur on timescales $\sim$$\beta^{-1}$or
and changes occur in distances $\sim\beta^{-1}\left(k_{B}T/m\right)$.
In terms of particles in the potential well we may safely assume that
they are in equilibrium with a concentration in the well that varies
as $w_{A}\: exp-m\psi/k_{B}T$ and a Maxwell distribution of particle
velocities $\phi(v)$ independent of $x$ given by 
\begin{equation}
\phi(v)=\frac{1}{\sqrt{2\pi k_{B}T/m}}exp-\frac{v^{2}}{2k_{B}T}\label{eq:-4}
\end{equation}
 The region beyond $C$ is such that a particle only responds to the
force which attracts it towards B, and there is no diffusion back
into A. The potential around B acts as a perfect sink for particles.
There is no fluid resistance to the particles motion and effectively
the particles motion is such that immediately beyond C refers only
to those particles traveling in the $+ve$ $x$ direction. So we can
write the current $j$ immediately beyond C as 
\begin{eqnarray}
j & = & w_{A}exp-m\psi/k_{B}T\;\int_{0}^{\infty}v\phi(v)dv\nonumber \\
 & = & w_{A}exp-m\psi/k_{B}T\;\left(\frac{k_{B}T}{2\pi m}\right)^{1/2}\label{eq:-5}
\end{eqnarray}
 So dividing by $N_{A}$ in Eq.(\ref{eq:-2}) gives $P$ 
\begin{equation}
P=j/N_{A}=\frac{\omega_{A}}{2\pi}exp-m\psi/k_{B}T\label{eq:-3}
\end{equation}

Unlike the rate constant based on Smoluchowski's advection diffusion
equation, the rate constant $P$ in Eq.(\ref{eq:-3}) does not depend
on $\omega_{C}$. The formula is based on what is called the transition
state method \cite{key-3}.

\section{Rate constant from a solution of the Fokker-Planck equation}

Chandrasekhar \cite{key-2} has evaluated the rate constant by solving
the Fokker-Planck equation for Brownian particles in a potential well
and obtained a solution that is valid for all $\beta$. The solution
necessarily contracts to the formulae previously derived for $\beta^{-1}\omega_{C}\ll1,\gg1$.
The same assumption is made that the height of the potential barrier
at C in Figure \ref{fig:Potential-wells} ,$Q\gg k_{B}T$. We shall
briefly outline how the formula is obtained and refer to Chandrasekhar
\cite{key-2} for the precise details. The Fokker-Planck equation
for the phase space density $W(v,x,t)$ for diffusion in a conservative
field of force $\ve K(\vecx)$ is given for 1D in space by 
\begin{equation}
\frac{\partial W}{\partial t}+v\frac{\partial W}{\partial x}+K\frac{\partial W}{\partial v}=\beta v\frac{\partial W}{\partial v}+\beta W+q\frac{\partial^{2}W}{\partial v^{2}}\label{eq:Fokker Planck Eqn}
\end{equation}
 where $K=-\partial\psi(x)/\partial x$ .The equilibrium solution
is 
\[
W(v,x)=Cexp\left[-m\left(v^{2}+2\psi(x)\right)/2k_{B}T\right]
\]
 for which $\psi(0)=0$ at A in Figure \ref{fig:Potential-wells}
and for future reference 
\begin{equation}
C=w_{A}\left(m/2\pi k_{B}T\right)^{1/2}\label{eq:-9}
\end{equation}
 As Chandrasekhar points out this cannot be the solution for all $x$
because there would be no diffusion of particles over the potential
barrier. For the case we are studying where $Q\gg k_{B}T$ we would
expect the solution to be close to the equilibrium solution at $A$.
A solution is therefore sought of the form 
\begin{equation}
W(v,x)=C\: F(v,x)xp\left[-m\left(v^{2}+2\psi(x)\right)/2k_{B}T\right]\label{eq:-6}
\end{equation}
 where $F(v)$is very near unity in the neighborhood $A$ at $x=0$and
for $x\gg x_{C}$ i.e. in the region of $B$. Chandrasekhar points
out that the most sensitive region to calculate the departure from
equilibrium in the neighborhood of $C$ where $\psi(x)$ has the form
given in Eq.(\ref{eq:potenial around C}) So substituting the form
for $W(v,x)$ in Eq.(\ref{eq:-6}) into the stationary Fokker-Planck
equation with $\psi(x)$ given by Eq.(\ref{eq:potenial around C})
of the Eq.(\ref{eq:Fokker Planck Eqn}) gives the following equation
for $F(v,x)$:

\begin{equation}
v\frac{\partial F}{\partial X}+\omega_{C}^{2}X\frac{\partial F}{\partial v}=q\frac{\partial^{2}F}{\partial v^{2}}-\beta v\frac{\partial F}{\partial v}\label{eq:eq.for F}
\end{equation}
 where we have written $F$ as a function of $X=x-x_{C}$ and $v$.
The solution of this equation satisfying the boundary conditions is
\begin{equation}
F(\xi)=\left(\frac{a-\beta}{2\pi q}\right)^{1/2}\int_{-\infty}^{\xi}exp\left[-(a-\beta)\xi^{2}/2q\right]d\xi\label{eq:-7}
\end{equation}
 where$\xi=v-aX$ and $a=\left[\left(\beta/2\right)^{2}+\omega_{C}^{2}\right]^{1/2}+\left(\beta/2\right)$.
The current at $C$ is given by 
\[
j=\int_{-\infty}^{\infty}W(v,X=0)vdv
\]
 which substituting the form for$W(v,x)$ based on Eq.(\ref{eq:-6}
and Eq.(\ref{eq:-7}) for $F(\xi$) gives 
\begin{equation}
j=C\left(\frac{k_{B}T}{m}\right)\left[(a-\beta)/a\right]^{1/2}e^{-mQ/k_{B}T}\label{eq:-8}
\end{equation}
 Using the formula in for $N_{A}$in Eq.(\ref{eq:-2} and the value
$C=$ the rate constant $P$ is given by 
\[
P=j/N_{A}=\left(\omega_{A}/2\pi\omega_{c}\right)\left\{ \left[\beta^{2}/4+\omega_{C}^{2}\right]^{1/2}-\beta/2.\right\} e^{-mQ/k_{B}T}
\]
 Note that for $\omega_{C}/\beta\gg1$, $P$ contract to the transition
state approximation Eq.\ref{eq:-3} and when $\omega_{C}/\beta\ll1$
$P$ contracts to the advection gradient diffusion approximation Eq.(\ref{eq:Smoluchowski rate constant})

\begin{wrapfigure}{R}{0.5\columnwidth}%
\includegraphics[scale=0.5]{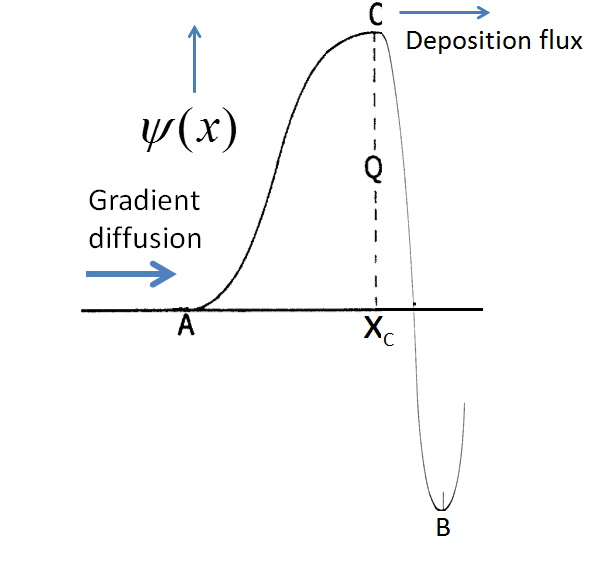} \caption{\label{fig:Potential Barrier for depositing particles}Potential Barrier
for particles depositing at a wall }
\end{wrapfigure}%

\section{Deposition velocity for escape of Brownian particles over potential
barriers.}

In the previous section we considered the case where all the particles
were contained in a potential well.The height of the potential barrier
$Q\gg k_{B}T/m$ the thermal energy /unit particle mass which meant
that most of the particles were located in the neighborhood of the
minimum potential at A and were in thermal equilibrium. We used the
number of particles in the well as a normalizing factor in the sense
that the flux of particles escaping from the well was directly proportional
to the number ofparticles in the well at any instant of time. So the
ratio $j/N_{A}$ was a constant in time and defined a rate constant
$P$ or equivalently the probability per unit time of a particle in
the potential well at A of escaping over the potential barrier at
C and into the potential well at B. $P^{-1}$ is therefore typical
of the lifetime of particles in the potential well around A.

We want now to consider a slightly different situation depicted in
Figure \ref{fig:Potential Barrier for depositing particles} corresponding
to particles being deposited at a wall at $B$ by diffusion but having
to overcome a potential barrier at C before reaching the wall. The
particles reach A from the left by Brownian diffusion from a region
where the external force acting on the particles is zero. We may assume
that in steady state the flux at A scales on the concentration $w_{A}$
so $j=k_{d}w_{A}$. The formula for the deposition velocity $k_{d}$
is straight forward to obtain since it is readily obtained from the
expression for $j$ we have obtained from the various cases we reconsidered
before (there is no need to divide by the number of particles in the
well to obtain the rate constant.) So from the expression for $j$
given in Eq.(\ref{eq:-8}) and substituting the value of $C$ in Eq.(\ref{eq:-9})
we have 
\begin{eqnarray}
k_{d} & =j/w_{A}= & \left(\frac{k_{B}T}{2\pi m}\right)^{1/2}\left[(a-\beta)/a\right]^{1/2}e^{-mQ/k_{B}T}\nonumber \\
 & = & \omega_{C}^{-1}\left(\frac{k_{B}T}{2\pi m}\right)^{1/2}\left(\left[\left(\beta/2\right)^{2}+\omega_{C}^{2}\right]^{1/2}-\beta/2\right)e^{-mQ/k_{B}T}\label{eq:-10}
\end{eqnarray}
 So for the value of $k_{d}$ based on the transition state method
for which ($\omega_{C}/\beta\gg1)$ we have from Eq.(\ref{eq:-10})
\[
k_{d}=\left(\frac{k_{B}T}{2\pi m}\right)^{1/2}e^{-mQ/k_{B}T}\;\;\left(\omega_{C}/\beta\right)\gg1
\]

and for $(\omega_{C}/\beta\ll1)$ (based on Smoluchowski's equation)
\begin{eqnarray*}
k_{d} & = & \beta/2\left(\left[1+\left(\frac{2\omega_{C}}{\beta}\right)^{2}\right]^{1/2}-1\right)\omega_{C}^{-1}\left(\frac{k_{B}T}{2\pi m}\right)^{1/2}e^{-mQ/k_{B}T}\;\;\\
 & = & \frac{\beta}{2\omega_{c}}\left[\frac{4\omega_{C}^{2}}{2\beta^{2}}\right]\left(\frac{k_{B}T}{2\pi m}\right)^{1/2}e^{-mQ/k_{B}T}\\
k_{d} & = & \left(\frac{\omega_{c}}{\beta}\right)\left(\frac{k_{B}T}{2\pi m}\right)^{1/2}e^{-mQ/k_{B}T}
\end{eqnarray*}

\end{document}